\documentclass[12pt]{article}
\usepackage{graphicx}
\usepackage{cite}
\def\be {\begin{equation}}
\def\ee {\end{equation}}
\def\beqy{\begin{eqnarray}}
\def\eeqy{\end{eqnarray}}
\def\jpsi{J/\psi}
\def\etac{\eta_c}
\def\as{\alpha_s}

\begin{document}

\title{Two Photon Width of $\eta_c$}

\author{Nicola Fabiano\\
\textsl{ Perugia University and INFN, via Pascoli I-06100, Perugia, Italy}\\
Giulia Pancheri\\
\textsl{INFN National Laboratories, P.O.Box 13, I00044 Frascati, Italy}}
\date{}

\maketitle

\begin{abstract}
We discuss the measured partial width of the pseudoscalar
charmonium state $\eta_c$  into two photons.
Predictions from potential models are examined and compared with
experimental values. Including radiative corrections, it is found
that present measurements are compatible both with a 
QCD type potential and with a static Coulomb potential. 
The latter is then used to give
an estimate on the $\eta_b$ decay into two photons.
Results for $\etac$ are also compared with those
from $\jpsi$ data through the NRQCD model.
\end{abstract}

\section{Introduction}

In this paper we
revisit the calculation of the two photon width of $\eta_c$, 
highlighting newest experimental 
results and updating the potential model calculation. This  allows for a
reliable estimate of the two photon width of $\eta_b$, which is been 
searched for  in $\gamma \gamma $ collisions \cite{armin1}. We shall see that
 the
expected two photon width of $\eta_b$ is within reach of the 
precision in the LEP data being analyzed.

The  charmonium spectrum has been  the basic testing grounds
for a variety of models for the interquark potential, ever since the
discovery of the $J/\psi$ in 1974 \cite{TING}. The experimental
scenario describing the  $c{\bar c}$ bound states is  close to
completion, with the observed higher excitation states
$^3P_0$, $^3P_1$  and spin 2 $^3P_2$ states \cite{E760} : decay 
widths into various
leptonic and hadronic states have been measured and compared with
potential models \cite{ROSNER,FRANZINES}. Most of this note is
dedicated to examine  the theoretical predictions for the electromagnetic
decay of the simplest
and lowest lying of all the charmonium states, i.e. the pseudoscalar 
$\eta_c$.  In Sect. 2 we shall compare the two photon decay width with 
 the leptonic width of the $J/\psi$, which has been  
measured with  higher precision \cite{PSI} and found to be 15\%
 higher than in previous measurements \cite{PDGPSI}. This implies 
that a number of potential models whose parameters had been determined
by the leptonic width of the $J/\psi$ may need some updating, and so
do some predictions from these models. Potential model predictions for
$\eta_c\to \gamma \gamma$  can be found in
Sect. 3, together with  a value for the two photon width of $\eta_b$ 
extracted from the Coulombic potential.
In Sect. 4 we show the predictions for $\eta_c$ decay widths, using
the  procedure introduced in~\cite{BBL} for the description of 
mesons made out of two non relativistic heavy quarks, by means of the Non
Relativistic Quantum Chromodynamics--NRQCD. 
In Sect. 5 we
 compare these different  determinations with the experimental value 
of the 
$\etac \to \gamma \gamma$ decay
width 
expanding some recent 
theoretical analyses on this subject
(see for instance~\cite{PENNINGTON},~\cite{CZARNECKI} and references therein).

 \section{Experimental values and relation to $J/\psi$ electromagnetic width}

\begin{figure}[t]
\begin{center}
\includegraphics[angle=0,width=0.7\textwidth]{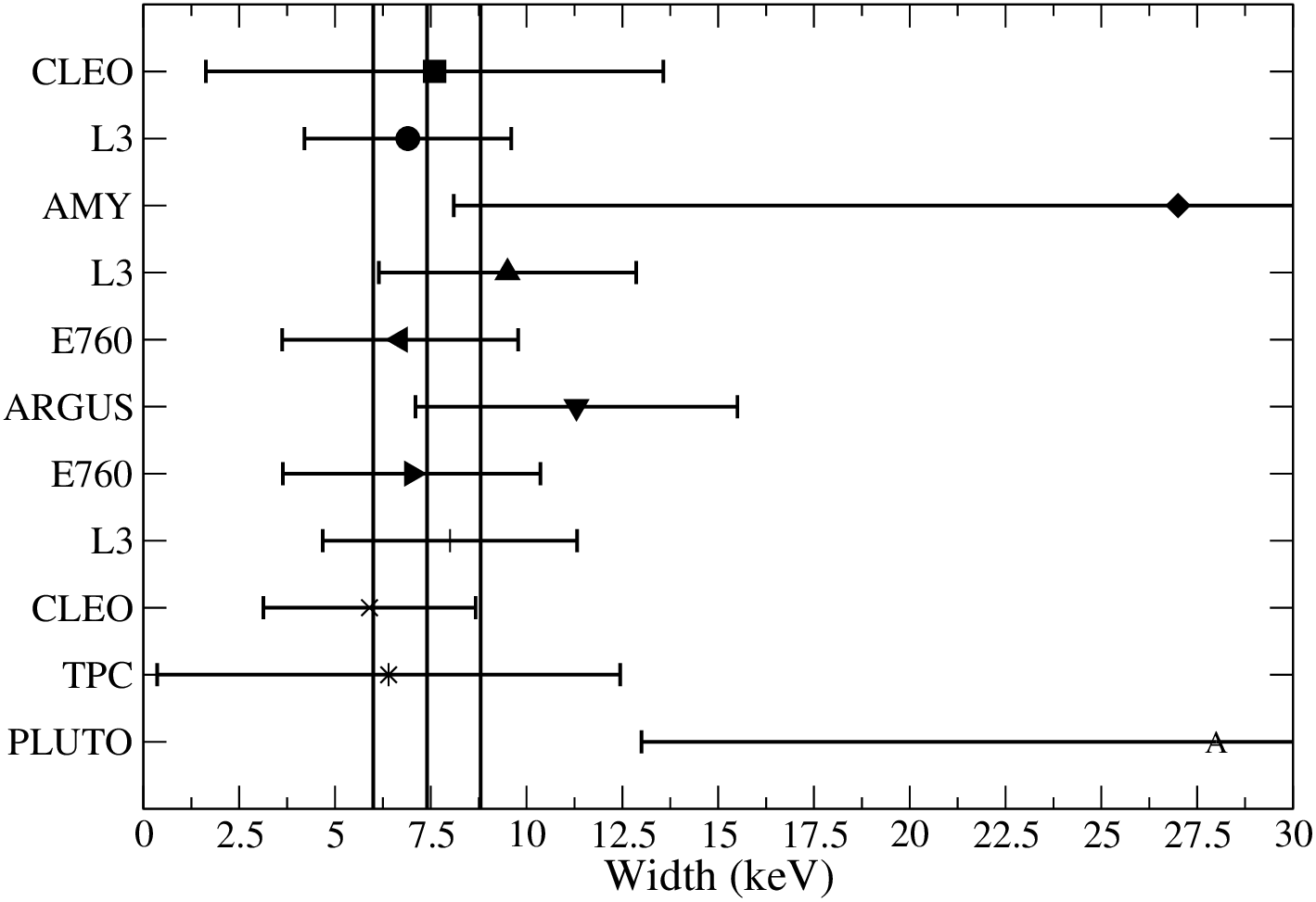} 
\caption{Experimental values of various measures of $\etac \to \gamma \gamma$
  \label{fig:expdata}}
\end{center}
\end{figure}

The first evidence of the $\eta_c$ state has been found in the inclusive 
photon
spectra of the $\psi'$ and $\jpsi$ decays~\cite{SLAC1},~\cite{SLAC2}.
Subsequently, through $\gamma \gamma$ collisions, the decay width of 
$\eta_c$ 
into two photons  has been measured in
different experiments. The
most recently  reported values for the radiative decay width  
are shown in fig.~(\ref{fig:expdata})~\cite{CLEO2000,AMY,L39909,E7601,
ARGUSDALLAS,E76094,L3,CLEO,TPC/2,PLUTO},
together with  the Particle Data Group average \cite{PDG}, which reads
\be
\Gamma_{exp}(\eta_c \to \gamma \gamma)=7.4 \pm 1.4  \textrm{ keV .}
\label{eq:etactogamgam}
\ee
In order to compare the experimental determinations with theoretical
expectations, we start with the two photon decay width of a pseudoscalar
quark-antiquark bound state \cite{VANROYEN}
with first order QCD corrections \cite{BARBIERI}, which can be written 
 as
\be
\Gamma(\eta_c\rightarrow \gamma\gamma)= \Gamma_B^P\left 
[ 1 + 
\frac{\alpha_s}{\pi} 
\left(\frac{\pi^2-20}{3} \right ) \right ]
 \label{eq:widsc1l} .
\ee
In eq.~(\ref{eq:widsc1l}),
  $\Gamma_B^P$ is  the Born decay width for a non-relativistic bound state
which can be calculated from  potential models.
A first theoretical estimate for this decay width can be obtained by 
comparing 
eq.~(\ref{eq:widsc1l})  with the expressions for the 
vector state $J/\psi$~\cite{MACKENZIE}, i.e.
\be
\Gamma (J/\psi\rightarrow e^+e^-)=\Gamma_B^V \left ( 1-
\frac{16}{3}\frac{\alpha_s}{\pi} \right )  .\label{eq:widvec1l}
\ee

The expressions in eqs.~(\ref{eq:widsc1l}) and~(\ref{eq:widvec1l}) can be 
used to estimate 
the radiative width of $\eta_c$ from the measured values of the
leptonic decay width of $J/\psi$, if one assumes  the same value
for  the wave
function at the origin $\psi(0)$,
 for both the pseudoscalar and the vector state. This is true up to errors
 of 
$\mathcal{O}(\alpha_s/
m_c^2)$
(see for instance \cite{CORNELL,EICHTEN}).

\begin{figure}[ht]
\begin{center}
\includegraphics*[angle=0,width=0.7\textwidth]{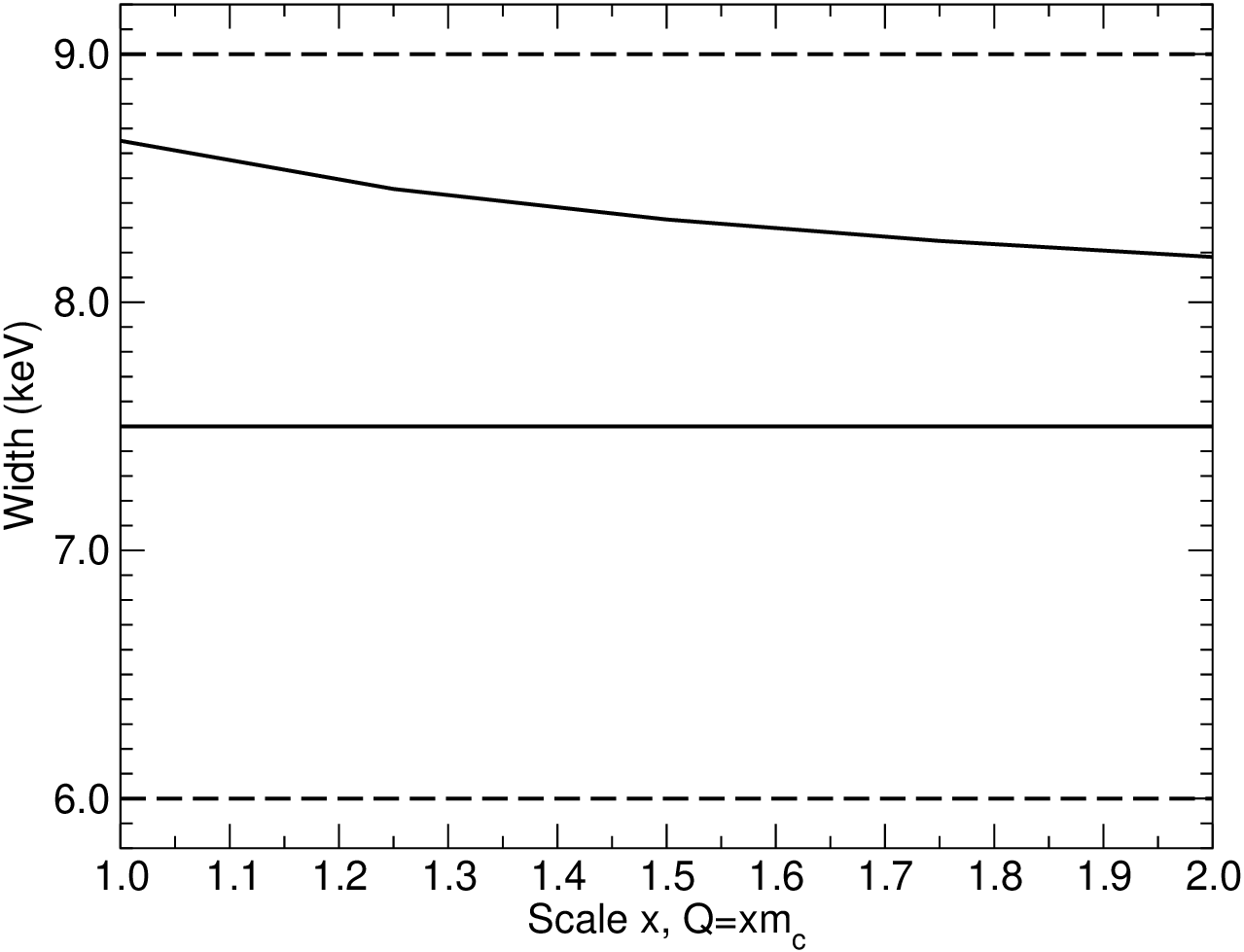} 
\caption{The dependence  of the 
$\eta_c$ decay width to $\gamma\gamma$ (in keV) is shown with 
respect to 
the scale chosen for $\as$~ in the radiative corrections. 
The horizontal lines represent the central 
experimental value (full line) and the    
(dashed lines) indetermination.
  \label{fig:asscale}}
\end{center}
\end{figure}

Taking the ratio between eqs.~(\ref{eq:widsc1l}) and (\ref{eq:widvec1l})
and expanding in $\as$, we obtain 
\be
\frac{\Gamma (\eta_c\rightarrow \gamma\gamma)}{\Gamma(J/\psi
\rightarrow e^+e^-)}\approx
\frac{4}{3} \frac{(1-3.38\as/\pi)}{(1-5.34\as/\pi)} = 
\frac{4}{3}  \left [ 1+1.96  \frac{\alpha_s}{\pi} 
+ \mathcal{O}(\alpha_s^2) \right ]
\label{eq:rappwid}
\ee
The correction can be computed  from the
two loop expression for $\as$
and the  value~\cite{PDG}
$\alpha_s(M_Z) = 0.118\pm 0.003 $. 
Using the  renormalization group equation to evaluate 
$\alpha_s(Q=2m_c=3.0\mbox{ GeV})=0.25 \pm 0.01$, and the  latest 
measurement    
\be
\Gamma_{exp} (J/\psi\rightarrow e^+e^-)= 5.26 \pm 0.37 \textrm{ keV}
\label{eq:expjpsiee}
\ee
one  obtains
\be
\Gamma(\eta_c\rightarrow \gamma\gamma) \pm
\Delta\Gamma(\eta_c\rightarrow \gamma\gamma) =8.18
\pm 0.57 \pm 0.04\ \ \textrm{keV}
\ee
where the first error comes from the uncertainty on the $\jpsi$ experimental
width, the second error from $\alpha_s$~.
This estimate agrees within $1 \sigma$ with the value given
in eq.~(\ref{eq:etactogamgam}).
Here we assumed the $\as$ scale to be $Q=2m_c=3.0\textrm{ GeV}$. This
choice is by no way unique, and in
fig.~(\ref{fig:asscale}) we show the dependence of the $\etac$ photonic 
width, evaluated  from eq.~(\ref{eq:expjpsiee}), upon  different values 
of the scale chosen for $\as$.

As one can see from fig.~(\ref{fig:asscale}) the experimental width value
is not sufficient to uniquely determine the a scale choice of $\as$~. 
We shall therefore include this
fluctuation in the indetermination due to radiative corrections.
\section{Potential models predictions for $\eta_c$ and $\eta_b$}
We present now the results one can obtain for
the absolute width, through the extraction of the wave
function at the origin from potential models.
For the calculation of the wavefunction we have used four different 
 models,
namely  the Cornell type potential \cite{CORNELL}
$V(r) = -\frac{k}{r} + \frac{r}{a^{2}} $
with parameters $a=2.34$ and $k=0.52$, the Richardson potential 
\cite{RICHARDSON}
$V_{R}(r) = -\frac{4}{3} \frac{12 \pi}{33-2N_{f}} \int 
\frac{d^{3}q}{(2\pi)^{3}} \frac{e^{iqr}}{q^{2}\log(1+q^{2}/
\Lambda^{2})}$
with $N_{f} = 3$ and $ \Lambda=398$ $MeV$, and 
the QCD inspired potential $V_{J}$  of Igi-Ono \cite{IGI,TYE}
\be
 V_{J}(r) = V_{AR}(r) + d r e^{-gr} + ar , \ \ \
 V _{AR}(r) = -\frac{4}{3} \frac{\alpha_{s}^{(2)}(r)}{r} \label{eq:igipot}
\ee 
with two different  parameter sets, corresponding to 
 $\Lambda_{\overline{MS}}=0.5 \  GeV$
and $\Lambda_{\overline{MS}}=0.3 \  GeV$ respectively \cite{IGI}. We also 
show the results from a 
Coulombic type potential with the QCD coupling $\alpha_s$ frozen
to a value of $r$ which corresponds to the Bohr radius of the
quarkonium system, obtained by solving the equation 
$r_B=3/(2m_c\alpha_s(r_B))$ ~\cite{NOI}. We stress that the scale of $\as$ 
occurring in the radiative corrections and the one of the Coulombic potential
are different\cite{FADIN}.

\begin{figure}[h]
\begin{center}
\includegraphics*[angle=0,width=0.7\textwidth]{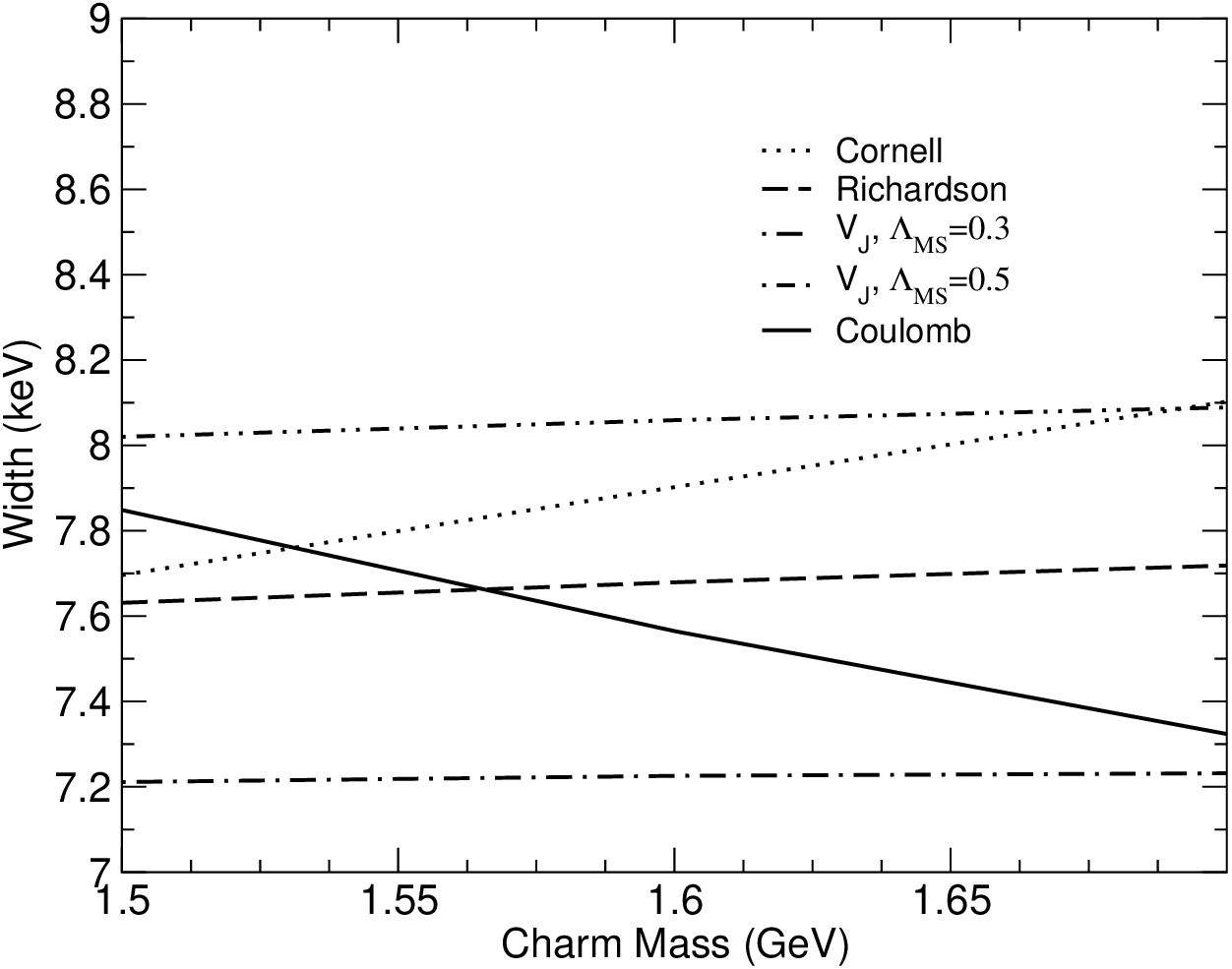}  
\caption{The dependence of $\eta_c$ decay width to $\gamma\gamma$ in 
\emph{keV} for  different
potential models is shown as a function of $m_c$~.   \label{fig:potentials}}
  \end{center}
\end{figure}
We show in fig.~(\ref{fig:potentials}) the predictions for the decay width 
from these  potential models with the correction from 
eq.~(\ref{eq:widsc1l}) at an $\as$ scale $Q=2m_c$~, observing 
that the calculated widths stay well within one standard deviation of the 
width value given by eq.~(\ref{eq:etactogamgam}).
For any given model, sources of error in this calculation arise from the
choice of scale in the radiative correction factor and 
the choice of the parameters. Including the
fluctuations of the results given by the different models, we can 
estimate a range of values for the  potential model predictions for 
the radiative decay width  $\Gamma(\etac \to \gamma \gamma)$,
namely

\be
\Gamma(\etac \to \gamma \gamma) =7.6 \pm 1.5 \textrm{ keV \mbox{       }.} 
\label{eq:potmodpred}
\ee

ALEPH has recently started to search for $\eta_b$ into two
photons\cite{ARMIN} and it is interesting to see whether the potential
models can predict  for this decay a value within experimental reach.
Predictions of course will be affected by the error due to  the parametric
dependence of the given potential model, an error which can be quite 
large since most of the parameters have been tuned with the charmonium
system. On the other hand, the Coulombic potential gives results for
the charmonium system in agreement with all the other models, and, at the
same time, is 
relatively free of
such parameter dependence.  With   only the scale of $\alpha_s$ in the wave
function to worry about, it can 
 be  used for a reliable 
 estimate. For this purpose we shall make use
of the expression in eq.~(\ref{eq:widsc1l}) where this time $e_q=1/3$ and 
$m_b=5.0 \textrm{ GeV}$. This gives the potential model prediction
\be
\Gamma(\eta_b\rightarrow \gamma\gamma)=0.50 \pm 0.03 \textrm{ keV} \; ,
\label{eq:etabresult}
\ee
where the error is associated to  different choices of $m_b$ values and
to the indetermination on $\as$ occurring in the radiative correction.
A check of this estimate can be given using the leptonic width of the
$\Upsilon$
and the expansion given in eq.~(\ref{eq:rappwid}).
To first order in $\as$ one obtains:
\be
\frac{\Gamma (\eta_b\rightarrow \gamma\gamma)}{\Gamma(\Upsilon
\rightarrow e^+e^-)}\approx
\frac{1}{3} \frac{(1-3.38\as/\pi)}{(1-5.34\as/\pi)} = 
\frac{1}{3}  \left [ 1+1.96  \frac{\alpha_s}{\pi} 
 + \mathcal{O}(\alpha_s^2) \right ]
 \label{eq:rapp-b}
\ee
which differs from eq.~(\ref{eq:rappwid}) only by a charge factor; using
the PDG average~\cite{PDG}
\be
\Gamma_{exp}(\Upsilon \to e^+e^-) = 1.32 \pm 0.05 \textrm{ keV}
\label{eq:expetab}
\ee
and assuming the wavefunctions of the two $b \overline{b}$ bound states to be 
equal we  have
\be
\Gamma(\eta_b \to \gamma \gamma) = 0.49 \pm 0.04 \textrm{ keV}
\label{eq:etab-xper}
\ee
in agreement with the Coulombic model prediction eq.~(\ref{eq:etabresult}).
For the radiative correction factor we  have used 
$\alpha_s(Q=2m_b=10 \textrm{ GeV})=0.18 \pm 0.01 $.
The associated error in eq.~(\ref{eq:etab-xper}) takes into account the 
indetermination on the experimental value eq.~(\ref{eq:expetab}) and
the one on $\as$~.

A more thorough discussion on the theoretical estimates of the
$\eta_b$ decay into two photons and comparison to some recent
calculations (see for instance~\cite{SCHULER})
will follow in a future publication~\cite{MEETAB}.

\section{Octet component model}

We will  present now another model which admits other components to the meson
decay beyond the one from the colour singlet picture (Bodwin, Braaten and
Lepage)~\cite{BBL}. 
NRQCD has been used to separate the short distance scale of 
annihilation from the nonperturbative contributions of long distance scale.
This model has been successfully used to
explain the larger than expected $\jpsi$ production at the
Tevatron and LEP.
According to BBL, in the octet model for quarkonium, the decay widths of
charmonium states are given by:

\beqy
\Gamma(J/\psi \to LH) &=&  
\frac{2\langle \jpsi | O_1(^3S_1) | \jpsi \rangle}
{m_c^2}  \frac{10\alpha_s^2(\pi^2-9)}{243} \times  \nonumber\\
\times \left [ \as \left ( 1-3.7 \frac{\alpha_s}{\pi} \right ) \right.  & + &  
 3.2  \alpha \left ( 1 - 6.7 \frac{\alpha_s}{\pi}
\left. \right ) \right ] - \nonumber \\
&-&\frac{2\langle \jpsi | P_1(^3S_1) | \jpsi \rangle}{m_c^4} 
\alpha_s^2 \left ( 2.0   \alpha + 0.6 \right )
\eeqy

\beqy
\Gamma(J/\psi \to e^+e^-) &=& \frac{8 \pi \alpha^2 \langle \jpsi 
| O_1(^3S_1) | \jpsi \rangle }{27 m_c^2} 
\left ( 1- \frac{16}{3}  \frac{\alpha_s}{\pi} \right )  \nonumber \\
&-&\frac{32\pi\alpha^2\langle\jpsi| P_1(^3S_1)|\jpsi \rangle}{81 m_c^4}
\eeqy

\beqy
\Gamma(\eta_c \to LH)& = &\frac{2 \pi \alpha_s^2 \langle \etac | O_1(^1S_0) |
 \etac \rangle }{9 m_c^2} 
\left [ 1 + \left ( \frac{143}{6} - \frac{31}{24}  \pi^2  \right ) \right.  \nonumber\\ 
\times  \left.\frac{\alpha_s}{\pi} \right ] & -& 
\frac{8 \pi \alpha_s^2 \langle \etac | P_1(^1S_0) | \etac\rangle}{27m_c^4}
\eeqy

\beqy
\Gamma(\eta_c \to \gamma \gamma) &=& \frac{32 \pi \alpha^2  \langle \etac 
| O_1(^1S_0) | \etac \rangle }{81 m_c^2} 
\left [ 1 + \left ( \frac{\pi^2}{4} - 5  \right ) \frac{4}{3}  
\frac{\alpha_s}{\pi} \right ] \nonumber \\
&-&\frac{128 \pi \alpha^2 \langle \etac | P_1(^1S_0) | \etac \rangle}{243m_c^4} 
\eeqy
There are four unknown long distance coefficients, which can be reduced to
two by means of the vacuum saturation approximation:
\be
G_1 \equiv \langle \jpsi | O_1(^3S_1) | \jpsi \rangle = \langle \etac 
| O_1(^1S_0) | \etac \rangle
\ee
\be
F_1 \equiv \langle \jpsi | P_1(^3S_1) | \jpsi \rangle = \langle \etac 
| P_1(^1S_0) | \etac \rangle
\ee
correct up to $\mathcal{O}(v^2)$, where ${\vec{v}}$ is  the quark
velocity inside the meson.
We use the $J/\psi$ experimental decay widths as input in order to determine
the long distance coefficients $G_1$ and $F_1$~. This result in turn
is used to compute the $\etac$ decay widths.

The BBL model gives the following decay widths of the $\etac$
meson:
\be
\Gamma(\eta_c\rightarrow \gamma\gamma)  =9.02 \pm 0.65 \pm 0.14 \textrm{ keV}
\ee
and
\be
\Gamma(\eta_c\rightarrow LH)  =14.38 \pm 1.07 \pm 1.43 \textrm{ MeV}
\ee
where the first error comes from the uncertainty on the $\jpsi$ experimental
width, the second error from $\alpha_s$~.
This results agree with experimental data within $1\sigma$, confirming the
applicability of the BBL model to the charm system. We leave to a future
publication the application of the BBL to the $b$--system.

\section{Comparison between models}

\begin{figure}[t]
\begin{center}
\includegraphics*[angle=0,width=0.7\textwidth]{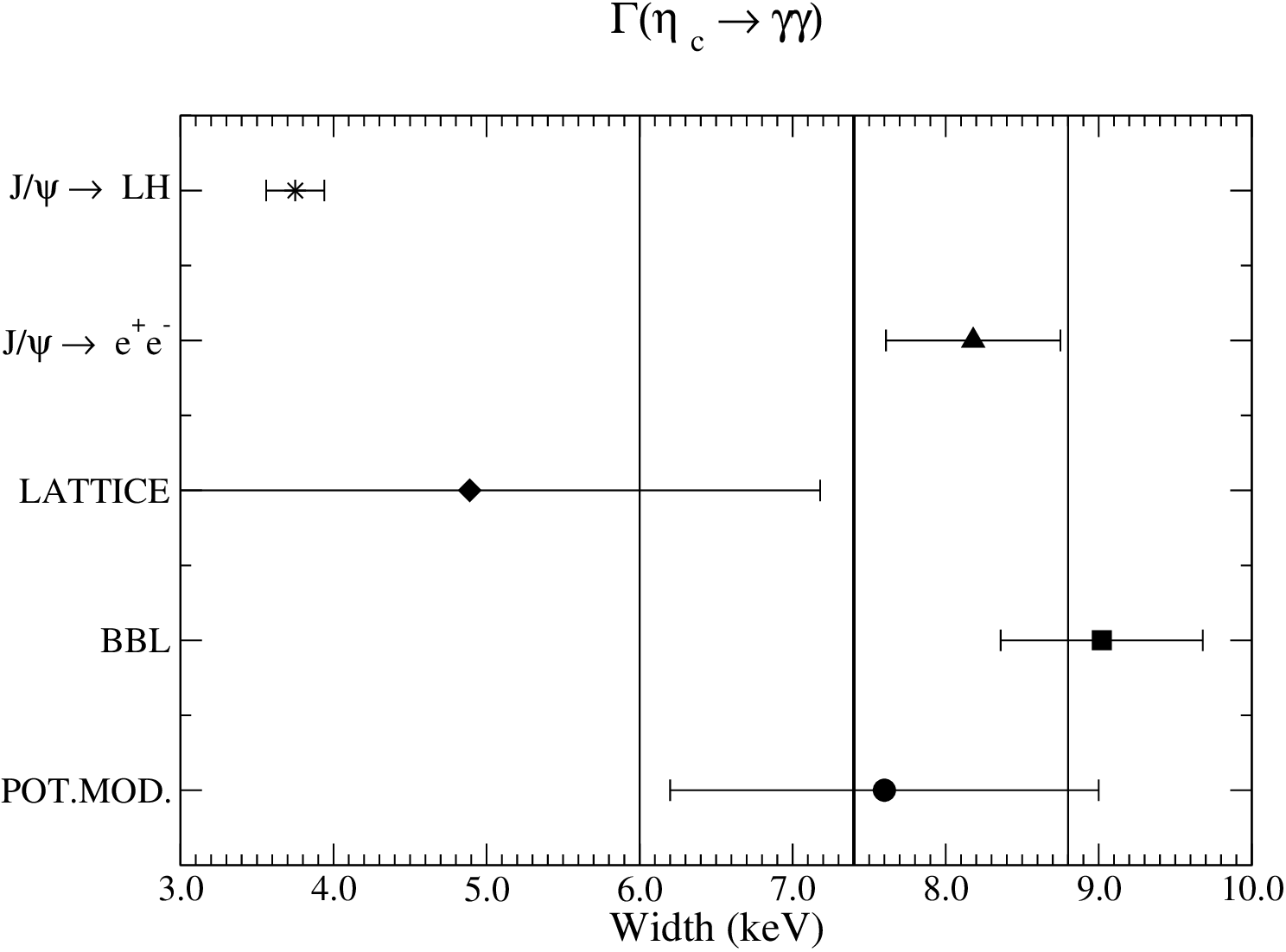}  
\caption{ The $\etac$ two photon width as calculated in
this paper using (starting from below)
Potential Models results, BBL model with input from $J/\psi$ 
decay data, Lattice evaluation of
$G_1$ and $F_1$ factors, Singlet picture with $G_1$ obtained from 
$J/\psi \to e^+e^-$ and $J/\psi \to LH$ processes respectively.
The scale has been enhanced with respect to fig.~2 to  evidentiate
the difference between various theoretical results. \label{fig:summary}}
\end{center}
\end{figure}

For comparison we present in fig.~(\ref{fig:summary})
a set of predictions coming from different
methods. We see that the theoretical results are in good
agreement which each other. 

Starting with   potential models, we see that the results are in excellent
agreement with the experimental world average taken from PDG. The advantage
of this method is that we are giving a prediction from first principles, 
without using any experimental input. The second evaluation, given by BBL 
using
the experimental values of the $\jpsi$ decay, is  off by 1$\sigma$ from
the central value. This is true also for the determination of the BBL model 
with nonperturbative long distance terms taken from
from the lattice calculation~\cite{LATTICE}, affected from a large error. 
The advantage of 
the latter is that its prediction, like the one from potential models, does
not make use of any experimental value. Next is the point given by the singlet
picture from the electromagnetic decay of the $\jpsi$,  in agreement
with the central value of the $\etac$. The last point is obtained also from 
the singlet picture with the $\jpsi$ decay into light hadrons, and is in
disagreement with the experimental measure.
 This is a long standing problem with some charmonium decay widths that
hasn't been resolved yet (see for instance~\cite{SETH} and references therein).

\section{Conclusions}
The $\Gamma(\etac \to \gamma \gamma)$ decay width prediction of the 
potential models considered gives the value $7.6 \pm 1.5 \textrm{ keV}$
which is consistent with 
the individual measurements and the world average\cite{PDG}. 
The Coulombic model is in agreement
with predictions from other models, and gives 
for the  $\eta_b \to \gamma \gamma$ decay width the estimate
$ 0.50 \pm 0.03 \textrm{ keV. }$
Predictions of the BBL model for the $\etac \to \gamma \gamma$
decay width are consistent with the experimental measurements, for both the long
distance terms $G_1$ and $F_1$ extracted from the $J/\psi$ experimental decay
widths and the one evaluated from lattice calculations.

\section*{Acnkowledgments}
G. P.  acknowledges partial support from EC Contract TMR98-0169.


\begin{thebibliography}{99}

\bibitem{armin1}A. B\"ohrer, {\it Bottom Production in Two Photon
Collisions at LEP}, DIS01, Bologna, Italy (2001), Proceedings Edi. by
G. Bruni, G. Iacobucci, R. Nania, World Sci., Singapore, 2001;
hep-ex/0106020;
\newline 
ALEPH Collaboration, {\it Search for $\gamma \gamma \rightarrow \eta_b$ in $e^+e^-$ collisions at LEP2} CERN-EP/2002-009; hep-ex/0202011.
\bibitem{TING}
J. J. Aubert et al., {\it Phys. Rev. Lett.} {\bf 33} (1974) 1404;
\newline
J. E. Augustin  et al., {\it Phys. Rev. Lett.} {\bf 33} (1974) 1406;
\newline
 G. S. Abrams  et al., {\it Phys. Rev. Lett.} {\bf 33} (1974) 1454.
\bibitem{E760}
T. A. Armstrong  et al.,  E760 Collaboration, {\it Phys. Rev.  Lett.} 
{\bf 70} (1993) 2988.
\bibitem{ROSNER}
A. K. Grant, J. L. Rosner and E. Rynes; {\it Phys. Rev. D} {\bf 47}
(1993) 1981.
\bibitem{FRANZINES}
P.L. Franzini, Juliet Lee--Franzini and P.L. Franzini,
{\it ``Fine structure of the P states in quarkonium and the spin dependent 
potential ''}, {\it LNF-}\mbox{$93/064$(P)}.
\bibitem{PSI}
J.Z. Bai  et al., {\it Phys. Lett. B} {\bf 355} (1995) 374; \\
S.Y. Hsueh and S. Palestini; {\it Phys. Rev. D} {\bf 45}
(1992) 2181.
\bibitem{PDGPSI}
A. M. Boyarski  et al., {\it Phys. Rev. Lett.} {\bf 34} (1975) 1357;
\newline
R. Baldini--Celio  et al., {\it Phys. Lett. B} {\bf 58} (1975) 471;
\newline
B. Esposito  et al., {\it Nuovo Cimento Lett.} {\bf 14} (1975) 73;
\newline
R. Brandelik  et al., {\it Z. Phys.} {\bf C1} (1979) 233.
\bibitem{BBL}
G.T. Bodwin, E. Braaten and G.P. Lepage, {\it Phys. Rev.~D} {\bf 51} (1995)
1125. 
\bibitem{PENNINGTON}
M.R. Pennington, {\it Nucl. Phys. B (Proc. Suppl.)} {\bf 82} (2000) 291.
\bibitem{CZARNECKI}
A. Czarnecki and K. Melnikov, \textit{Phys. Lett. B} \textbf{519} (2001) 212.
\bibitem{SLAC1}
T. Himel et al., \textit{Phys. Rev. Lett.} \textbf{45} (1980) 1146.
\bibitem{SLAC2}
R. Partridge et al., \textit{Phys. Rev. Lett.} \textbf{45} (1980) 1150.
\bibitem{CLEO2000}
G.~Brandenburg et al., {\it Phys. Rev. Lett.} {\bf 85} (2000) 3095.
\bibitem{AMY}
The AMY collaboration, M.~Shirai et al., 
 {\it Phys. Lett.~B} {\bf 424} (1998) 405.
\bibitem{L39909}
The L3 Collaboration, M. Acciarri et al., 
{\it Phys. Lett.~B} {\bf 461} (1999) 155.
\bibitem{E7601}
T. A. Armstrong  et al., E760 Collaboration,  {\it Phys. Rev. D} 
{\bf 52} (1995) 4839.
\bibitem{ARGUSDALLAS}
 H. Albrecht  et al., {\it Phys. Lett. B} {\bf 338} (1994) 390.
\bibitem{E76094}
D. Morgan, M. R. Pennington and M. R. Whalley, {\it J. Phys. G: Nucl. Part.
 Phys.} {\bf 20} (1994) A1-A147.
  \bibitem{L3}
L3 Collaboration, {\it Phys. Lett. B} {\bf 318} (1993) 575.
\bibitem{CLEO} 
CLEO Collaboration, W. Y Chen  et al., {\it Phys. Lett. B} {\bf 243} 
(1990) 169.
\bibitem{TPC/2}
TPC/2$\gamma$ Collaboration, {\it Phys. Rev. Lett.} {\bf 60} (1988) 2355.
\bibitem{PLUTO}
C.~Berger et al., {\it Phys. Lett. B} {\bf 167} (1986) 120.
\bibitem{PDG}
Review of Particle Properties, D.E.~Groom et al., 
{\it Euro. Phys. Journ.} {\bf C15} (2000) 1; \\http://pdg.lbl.gov/.
\bibitem{VANROYEN}
R.Van Royen and V.Weisskopf, {\it Nuovo Cimento} {\bf 50A} (1967) 617.
\bibitem{BARBIERI}
R. Barbieri, G. Curci, E. d'Emilio and R. Remiddi
  {\it Nucl. Phys. B} {\bf 154} (1979) 535.
\bibitem{MACKENZIE}
P. Mackenzie and G. Lepage,  {\it Phys. Rev. Lett.} {\bf 47} (1981) 1244.
\bibitem{CORNELL}
E. Eichten, K. Gottfried, T. Kinoshita, K. D. Lane and T. M. Yan,
 {\it Phys. Rev. D} {\bf 21} (1980) 203.
\bibitem{EICHTEN}
E. Eichten and F. Feinberg, {\it Phys. Rev. D} {\bf 23} (1981) 2724.
\bibitem{2LOOPS}
W.A. Bardeen, A.J. Buras, D.W. Duke and T. Muta, \textit{Phys. Rev. D} 
\textbf{18} (1978) 3998; \\
W.J. Marciano, \textit{Phys. Rev. D} \textbf{29} (1984) 580. 
\bibitem{RICHARDSON}
J.L. Richardson, {\it Phys. Lett. B} {\bf 82} (1979) 272.
\bibitem{IGI}
J. H. K\"uhn and S. Ono, {\it Zeit. Phys.} {\bf C21} (1984) 385; \newline
K. Igi and S. Ono, {\it Phys. Rev. D} {\bf 33} (1986) 3349.
\bibitem{TYE}
W. Buchmuller and S. H. H. Tye, {\it Phys. Rev. D} {\bf 24} (1981) 132.
\bibitem{NOI}
N. Fabiano, A. Grau and G. Pancheri, {\it Phys. Rev. D} {\bf 50} (1994) 
3173;  {\it Nuovo Cimento A, Vol 107} (1994).
\bibitem{FADIN}
V.S. Fadin, V.A. Khoze, \textit{JETP Lett.} \textbf {46} (1987) 525; \newline
V.S. Fadin, V.A. Khoze, \textit{Yad. Fiz.} \textbf{48} (1988) 487.
\bibitem{ARMIN}
A. B\"ohrer, hep-ph/0110030, 
Talk presented at the International Conference on 
The Structure and Interactions of the Photon ``Photon 2001'', 
September $2^{nd}-7^{th}$ 2001, Ascona, Switzerland.
\bibitem{SCHULER}
G.A.~Schuler, F.A.~Berends and R.~van~Gulik, \textit{Nucl. Phys. B} \textbf{523}
(1998) 423.
\bibitem{MEETAB}
N. Fabiano, in preparation.
\bibitem{LATTICE}
G.T. Bodwin, D.K. Sinclair and S. Kim, {\it Int. J. Mod. Phys. A}
 {\bf 12} (1997) 4019.
\bibitem{SETH}
K.K. Seth, \textit{Nucl. Phys. B (Proc. Suppl.)} \textbf{71} (1999) 413.

\end{thebibliography}
\end{document}